\documentclass{bioinfo}
\usepackage{soul,color}
\copyrightyear{2008}
\pubyear{2008}

\begin{document}
\firstpage{1}

\title[$DIYABC$: a program for inferring population history]{Inferring population history with $DIYABC$: a user-friendly approach to Approximate Bayesian Computation}
\author[Cornuet \textit{et~al.}]{Jean-Marie Cornuet\,$^{1,2}$\footnote{to whom correspondence should be addressed}, Filipe Santos\,$^2$, Mark A. Beaumont\,$^3$, Christian P. Robert\,$^4$, Jean-Michel Marin\,$^5$, David J. Balding\,$^1$, Thomas Guillemaud\,$^6$ and Arnaud Estoup\,$^2$}
\address{$^1$Department of Epidemiology and Public Health, Imperial College, St Mary's Campus, Norfolk Place, London W2 1PG, U.K.\\
$^{2}$Centre de Biologie et de Gestion des Populations, INRA, Campus International de Baillarguet, CS 30016 34988 Montferrier-sur-Lez, France\\
$^3$School of Biological Sciences, Lyle Building, The University of Reading Whiteknights, Reading RG6 6AS, UK\\
$^4$CEREMADE, Universit\'e Paris-Dauphine, Place Delattre de Tassigny, 75775 Paris cedex 16, France\\
$^5$INRIA Saclay, Projet select, Universit\'e Paris-Sud, Laboratoire de Math\'ematiques (B\^at. 425), 91400 Orsay, France\\
$^6$UMR 1301 I.B.S.V. INRA-UNSA-CNRS. 400 Route des Chappes. BP 167 - 06903 Sophia Antipolis cedex. France}
\history{Received on XXXXX; revised on XXXXX; accepted on XXXXX}

\editor{Associate Editor: XXXXXXX}

\maketitle

\begin{abstract}

\section{Summary:}
Genetic data obtained on population samples convey information about their evolutionary history. Inference methods can extract part of this information but they require sophisticated statistical techniques that have been made available to the biologist community (through computer programs) only for simple and standard situations typically involving a small number of samples.  We propose here a computer program ($DIYABC$)  for inference based on Approximate Bayesian Computation (ABC),  in which scenarios can be customized by the user to fit many complex situations involving any number of populations and samples. Such scenarios involve any combination of population divergences, admixtures and population size changes. $DIYABC$ can be used to compare competing scenarios, estimate parameters for one or more scenarios, and compute bias and precision measures for a given scenario and known values of parameters (the current version applies to unlinked microsatellite data). This article describes key methods used in the program and provides its main features. The analysis of one simulated and one real data set, both with complex evolutionary scenarios, illustrates the main possibilities of $DIYABC$.
\section{Availability:}
The software $DIYABC$ is freely available at \href{http://www.montpellier.inra.fr/CBGP/diyabc}{http://www.montpellier.inra.fr/CBGP/diyabc}.
\section{Supplementary material:} Supplementary data are also available at \href{http://www.montpellier.inra.fr/CBGP/diyabc}{http://www.montpellier.inra.fr/CBGP/diyabc}
\section{Contact:} \href{j.cornuet@imperial.ac.uk}{j.cornuet@imperial.ac.uk}
\end{abstract}

\section{Introduction}

Until now, most literature and software about inference in population genetics concern simple standard evolutionary scenarios: a single population \citep{GT1994, SD2000, B1999}, two populations exchanging genes \citep{HN2004, DG2004} or not \citep{HST2007} or three populations in the classic admixture scheme \citep{W2003, Ex2005}. The main exception to our knowledge is the computer program $BATWING$ \citep{WB2003} which considers a whole family of scenarios in which an ancestral population splits into as many subpopulations as needed. However, in practice, population geneticists collect  and analyse samples which rarely correspond to one of these standard scenarios. If they want to apply methods developed in the literature and for which computer programs are available, they have to select subsets of samples (to fit these standard situations), at the price of lowering the power of the analysis. The other solution is to develop their own software, which requires specific skills or the right collaborators. Rare examples of inference in non standard scenarios can be found in \citet{OB1998} including 3 populations and two successive divergences, or \citet{EB2004} (10 populations that sequentially diverged with initial bottlenecks and exchanging migrants with neighbouring populations).\

Inference in complex evolutionary scenarios can be performed in various ways, but all are based on the genealogical tree of sampled genes and coalescence theory. A first approach used in programs such as $IM$ \citep{HN2004} or $BATWING$ consists of starting from a gene genealogy compatible with the observed data and exploring the parameter and genealogy space through MCMC algorithms. One difficulty with this approach is to be sure that the MCMC has converged, because of the huge dimension of the parameter space. With a  complex scenario, the difficulty is increased. Also, although not impossible, it seems quite challenging to write a program that would deal with very different scenarios. A second approach pioneered by \citet{B2003} consists in combining MCMC exploration of the scenario parameter space with an Importance Sampling (IS) based estimation of the likelihood. The strength of this approach is that the low number of parameters ensures a (relatively) fast convergence of the MCMC. Its weakness is that the likelihood is only approximated through IS, sometimes resulting in poor acceptance rates.\

When dealing with complex situations, the two previous approaches raise difficulties which mainly stem in the computation of the likelihood. Consequently, a line of research including  the works of \citet{TB1997}, \citet{WH1998}, \citet{P1999} and \citet{MM2003} developed a new approach termed Approximate Bayesian Computation (or ABC) by \citet{B2002}. In this approach, the likelihood criterion is replaced by a similarity criterion between simulated and observed data sets, similarity usually measured by a distance between summary statistics computed on both data sets. Among examples of inference in complex scenarios given above, all but one (the simplest) have used this approach, showing that it can indeed solve complex problems.\

 The ABC approach presents two additional features that can be of interest for experimental biologist. One characteristic, already noted by \citet{Ex2005}, is the possibility to assess the bias and precision of estimates for simulated data sets produced with known values of parameters with little extra computational cost. To get the same information with likelihood-based methods would require a huge amount of additional computation whereas, with ABC, the largest proportion of computation used for estimating parameters can be recycled in a bias/precision analysis. The second feature is  the simple way by which the posterior probability of different scenarios applied to the same data set can be estimated \citep[e.g.][]{ME2005,PC2007}.\

In its current state, the ABC approach remains inaccessible to most biologists because there is not yet a simple software solution. Therefore, we developed the program $DIYABC$ that performs ABC analyses on complex scenarios, i.e. which include any number of populations and samples (samples possibly taken at different times), with populations related by divergence and/or admixture events and possibly experiencing changes of population size. The current version is restricted to unlinked microsatellite data. In this article, we describe the rationale for some methods involved in the program. Then we give the main features of $DIYABC$ and we provide two complete example analyses performed with this program to illustrate its possibilities.  

\section{Key methods involved in DIYABC}

Inference about the posterior distribution of parameters in an ABC analysis is usually performed in three steps (see Figure S1 in Supplementary material). The first one is a simulation step in which a very large table (the \emph{reference table}) is produced and recorded. Each row corresponds to a simulated data set and contains the parameter values used to simulate the data set and summary statistics computed on the simulated data set. Parameter values are drawn from prior distributions. Using these parameter values, genetic data are simulated as explained in the next section. The summary statistics correspond to those traditionally used by population geneticists to characterize the genetic diversity within and among samples (e.g. number of alleles, genic diversity and genetic distances). The idea is to extract maximum genetic information from the data, admitting that exhaustivity or sufficiency are generally out of reach. The simulation step is generally the most time-consuming step since the number of simulated data sets can reach several millions. The second step is  a rejection step. Euclidian distances between each simulated and the observed data set in the space of summary statistics are computed and only the simulated data sets closest to the observed data set are retained. The parameter values used to simulate these selected data sets provide a sample of parameter values approximately distributed according to their own posterior distribution. \citet{B2002} have shown that a local linear regression (third step  = estimation step)  provides a better approximation of the posterior distribution.\ 

This synoptic of ABC is well established and we now concentrate on more specific issues that are implemented in $DIYABC$.
\subsection{Simulating genetic data in complex scenarios}
Thanks to coalescence theory, it has become easy to simulate data sets by a two steps procedure. The first step consists of building a genealogical tree of sampled genes according to rather simple rules provided by coalescence theory (see below). The second step consists of attributing allelic states to all the nodes of the genealogy, starting from the common ancestor and simulating mutations according to the mutation model of the genetic markers. In a complex scenario, only the first step needs special attention and we will concentrate on it now.\

In a single isolated population of constant effective size, the genealogical tree of a sample of genes is simulated backward in time: starting from the time of sampling, the gene lineages are merged (coalesced) at times that are drawn from an exponential distribution with rate $j(j-1)/4N_e$, when there are $j$ distinct lineages and the (diploid) effective population size is $N_e$. The genealogical tree is completed when there remains a single lineage. \

Consider now two isolated populations (effective population sizes $N_1$ and $N_2$ respectively) that diverged $t_d$ generations before their common sampling time. Since the two populations do not exchange genes, lineages within each population will coalesce independently. Coalescence simulation will stop either when there remains a single lineage or when the simulated time is beyond the divergence (looking back in time). In the latter case, the coalescence event is simply discarded. At generation $t_d$, the remaining lineages are simply pooled and will coalesce in the ancestral population. Because of the memoryless property of the exponential distribution, the time to the first coalescence in the ancestral population is independent of the times of the last coalescence in each daughter population and can be simulated as in the single isolated population above. Again, the genealogical tree is completed when there remains a single lineage in the ancestral population. Note that the two populations need not be sampled at the same generation since this has no bearing on the simulation process.\

Consider now the classic admixture scenario with  one admixed and two parental populations, as in  Figure 1 in \citet{Ex2005}. Simulating the complete genealogical tree can be achieved with the following steps: i) coalesce gene lineages in each population independently until reaching admixture time,
 ii) distribute remaining lineages of the admixed population among the two parental populations, each with a Bernoulli draw with probability equal to the admixture rate,  iii) coalesce gene lineages in the two parental populations until reaching their divergence time, iv) pool the remaining gene lineages of the two parental populations and place them into the ancestral population and v) coalesce gene lineages in the ancestral population. \
 
  We first note the modular form of this algorithm which involves only three modules:
 \begin{enumerate}
\item a module that performs coalescences in an isolated constant size population between two given times,
\item a module that pools gene lineages from two populations (for divergence),
\item a module that splits gene lineages from the admixed population between two parental populations (for admixture). 
\end{enumerate}

We also note that the last two modules are quite simple and that the first one might be extended to include population size variations.\ 

We have introduced a fourth module that proves useful in many instances. It performs the (simple) task of adding a gene sample to a population at a given generation. The interest of this module is to allow for multiple samples of the same population taken at different generations. By combining the aforementioned four modules, it is possible to simulate genetic data involving any number of populations according to a scenario that can include divergence, admixture events as well as population size variations. In addition, populations can be sampled more than once at different times. Compared to our previous definition of complex scenarios, the only restriction so far concerns the absence of migrations among populations. If migrations have to be taken into account, coalescences in two (or more) populations exchanging migrants are no longer independent and should be treated in the same module. Such a module would require to consider simultaneously two kind of events, coalescences of lineages within population and migrations of gene lineages from one population to another. In the current stage of $DIYABC$, this has not yet been achieved. \

\subsection{Two ways of simulating colescence events}
Simulating coalescences can be performed in two ways. The most traditional way is based on the usually fulfilled assumption that the effective population size is large enough so that the probability of coalescence is small and hence that the probability that two or more coalescences occur at the same generation is low enough so that it can be neglected \citep[e.g.][]{No2007}. Time is then considered as a continuous variable in computations. The corresponding algorithm, called here the \emph{continuous time} (CT) algorithm, consists in drawing first times between two successive coalescence events and then drawing 2 lineages at random at each coalescence event.\

However, in practice, population size can be so small (e.g. during a bottleneck) that multiple coalescences at the same generation become common, including with the same parental gene (producing multifurcating trees). Simulating gene genealogies with multiple coalescences is possible, \citep[e.g.][]{LE2004}. In effect, lineages are reconstructed one generation at a time: lineages existing at generation $g$ are given a random number drawn in $U[1,2N_e]$ and lineages with the same number coalesce together. The latter is termed here the \emph{generation by generation} (GbG) algorithm. \

The CT algorithm is much faster in most cases and is used in most softwares, but in some circumstances, the approximation becomes unacceptable. The solution taken in $DIYABC$ is to swap between the two algorithms according to a criterion based on the effective population size, the time during which the effective size keeps its value, and the number of lineages at the start of the module. The criterion is such that the generation per generation (GbG) algorithm is taken whenever it is faster (this occurs when the effective size is very small) or when the continuous time (CT) algorithm overestimates by more than 5\% on average the number of lineages at the end of the module.\

A specific comparison study has been performed to establish this criterion. For different time periods counted in number of generations ($g$), effective population sizes ($N_e$) and number of entering lineages ($n_{el}$), coalescences have been simulated according to each algorithm 10,000 times and the average number of remaining lineages at the end of the period have been recorded as well as the average computation duration of each algorithm. Our results (Figure S2) show that the following rules optimize computation speed while keeping the relative bias in coalescence rates under the 5\% threshold:\

if ($1<g\leq30$) do CT if  $n_{el}/N_e<0.0031g^2-0.053g+0.7197$ else do GbG\

if ($30<g\leq100$) do CT if $n_{el}/N_e<0.033g+1.7$ else do GbG\

if ($100<g$) do CT if $n_{el}/N_e<5$ else do GbG\

\subsection{Comparing scenarios}
Using ABC to compare different scenarios and infer their posterior probability has been performed in two ways in the literature. Starting with a reference table containing parameters and summary statistics obtained with the different scenarios to be compared (or pooling reference tables, each obtained with a given scenario), data sets are ordered by increasing distance to the observed data set. A first method (termed hereafter the \emph{direct} approach) is to take as an estimate of the posterior probability of a scenario the proportion of data sets obtained with this scenario in the $n_\delta$ closest data sets \citep{ME2005, PC2007}. The value of $n_\delta$ is arbitrary and unless the results are quite clear cut, the estimated posterior probability may vary with $n_\delta$. \

Following the same rationale that introduced the local linear regression in the estimation of posterior distributions for parameters \citep{B2002}, we perform a weighted polychotomous logistic regression to estimate the posterior probability of scenarios, termed hereafter the \emph{logistic} approach \citep[see also][]{FR2007,B2008}. In the estimation of parameters, a linear regression is performed with dependent variable the parameter and predictors the differences between the observed and simulated statistics. This linear regression is local at the point (in the predictor space) corresponding to the observed data set, using an Epanechnikov kernel based on the distance between observed and simulated summary statistics (see formula (5) in Beaumont \emph{et al.}, 2002). Parameters values are then replaced by their estimates at that point in the regression. \

Keeping the differences between observed and simulated statistics as the predictor variables in the regression, we consider now the posterior probability of scenarios as the dependent variable. Because of the nature of the "parameter", an indicator of the scenario number, a  $logit$ link function is applied to the regression. The local aspect of the regression is obtained by taking the same weights as in the linear adjustment of parameter values as described in \citet{B2002}. Confidence intervals for the posterior probabilities of scenarios are computed through the limiting distribution of the maximum likelihood estimators. See Annex 1 in Supplementary material for a detailed explanation.\
\subsection{Quantifying confidence in parameter estimations on simulated test data sets}

In order to measure bias and precision, we need to simulate data sets (i.e. test data sets) with known values of parameters and compare estimates with their true values. In the ABC estimation procedure, the most time-consuming task is to produce a large enough reference table. However, when such a reference table has been produced, e.g. for the analysis of a real data set, it can also be used to quantify bias and precision on test data sets as well.\

Measuring bias is straightforward, but precision can be assessed with different measures.  In $DIYABC$, the latter include the relative square root of the mean square error, the relative square root of the mean integrated square error, the relative mean absolute deviation, the 95 and 50\% coverages and the factor 2. See Annex 2 in Supplementary material for more details.

\section{DIYABC: a computer program for population biologists}
\subsection{Main features}
$DIYABC$ is a program that performs ABC inference on population genetic data. In its current state, the data are genotypes at microsatellite loci of samples of diploid individuals (missing data are allowed). The inference bears on the evolutionary history of the sampled populations by quantifying the relative support of data to possible scenarios and by estimating posterior densities of associated parameters. $DIYABC$ is a program written in Delphi running under a 32-bit Windows operating system (e.g. Windows XP) and it has a user-friendly graphical interface.\

The program accepts complex evolutionary scenarios involving any number of populations and samples.   Scenarios can include any number of the following timed events: stepwise change of effective population size, population divergence and admixture. They can also include unsampled as well as serially sampled populations as in \citet{B2003}. The main restriction regarding scenario complexity is the absence of migrations between populations.\

Since the program has been written for microsatellite data, it proposes two mutation models, namely the stepwise mutation model (SMM) and the generalized stepwise mutation (GSM) model \citep{E2002}. Note that the same mutation model has to be applied to all microsatellite loci, but these may have different values of mutation parameters.\

The historico-demographic parameters of scenarios may be of three types: effective sizes, times of events (in generations) and admixture rates. Marker parameters are mutation rates and the coefficient of the geometric distribution (under the GSM only). The program can also estimate composite parameters such as $\theta=4N_e\mu$ and $\tau=t\mu$, with $N_e$ being the diploid effective population size, $t$ the time of an event and $\mu$ the mean mutation rate. Prior distributions are defined for original parameters and those for composite parameters are obtained via an assumption of independence of their component prior densities. Priors for historico-demographic parameters can be chosen among four common distributions: Uniform, Log-uniform, Normal and Log-normal. Users can set minimum and maximum (for all distributions) and mean and  
standard deviation (for Normal and Log-normal). In addition, priors can be modified by setting binary conditions ($>$, $<$, $\geq$ and $\leq$) on pairs of parameters of the same category (two effectives sizes or two times of event). This is especially useful to control the relative times of events when these are parameters of the scenario. For priors of mutation parameters, only the Uniform and the Gamma distributions are considered, but hierarchical schemes are possible, with a mean mutation rate or coefficient P (of the geometric distribution in the GSM) drawn from a given prior and individual loci parameter values drawn from a gamma distribution around the mean.\
  
Available summary statistics are usual population genetic statistics averaged over loci: e.g. mean number of alleles, mean genic diversity, $F$st, ($\delta\mu)^2$, admixture rates ... \

Regarding ABC computations, the program can i) create a reference table or append values to an existing table, ii) compute the posterior probability of different scenarios, iii) estimate the posterior distributions of original and/or composite parameters for one or more scenarios and iv) compute bias and precision for a given scenario and given values of parameters . 
Finally, the program can be used simply to simulate data sets in the popular $Genepop$ format \citep{RR1995}.
 
\subsection{Two examples of analysis with $DIYABC$}
\subsubsection{Illustration on a simulated data set}
In order to illustrate the capabilities of $DIYABC$, we take first an example based on a data set simulated according to a complex scenario including three splits and two admixture events (scenario 1 in Figure 1). The scenario includes six populations: two of them have been sampled at time 0, the third one at time 2 and the fourth one at time 4, the last two have not been sampled. Each population sample includes 30 diploid individuals and data are simulated at 10 microsatellite loci. This scenario is not purely theoretical as it could be applied for instance to European populations of honeybees in which the Italian populations (\emph{Apis mellifera ligustica} result from an ancient admixture between two evolutionary branches \citep{Fr2000} that would correspond here to population samples 1 and 4. Furthermore, in the last 50 years, Italian bees have been widely exported and sample 2 could well correspond to a population of a parental branch that has been recently introgressed by Italian queens. This example also stresses the ability of $DIYABC$ to distinguish two events that are confounded in the usual admixture scheme: the admixture event itself and the time at which the real parental populations in the admixture diverged from the population taken as "parental".\

Our ABC analysis will address the following questions: 1) Suppose that we are not sure that the scenario having produced our example data set does include a double admixture and that we want to challenge this double admixture scenario with two simpler scenarios, one with a single admixture (scenario 2 in Figure 1) and the other with no admixture at all (scenario 3). What is the posterior probability of these three scenarios, given identical prior probabilities ? 2) What are the posterior distributions of parameters, given that the right scenario is known ? and 3) What confidence can we have in the posterior probabilities of scenarios and posterior distributions of parameters ?
  
First, a reference table is built up. Using different screens of $DIYABC$, i) the three scenarios are coded and prior distributions of parameters are defined (Figure S3), ii) based on previous studies (e.g. Dib \emph{et al.}, 1994 in Pascual \emph{et al.}, 2007), the Generalized Stepwise Mutation model is selected and prior distributions of mutation parameters are defined (Figure S4), iii) motif sizes and allele ranges of loci are set (Figure S5) and iv) summary statistics are selected (Figure S6). After some hours, a reference table with 6 million simulated data sets (i.e. 2 million per scenario) is produced.\

\begin{figure}[h]
\centerline{\includegraphics[scale=0.1]{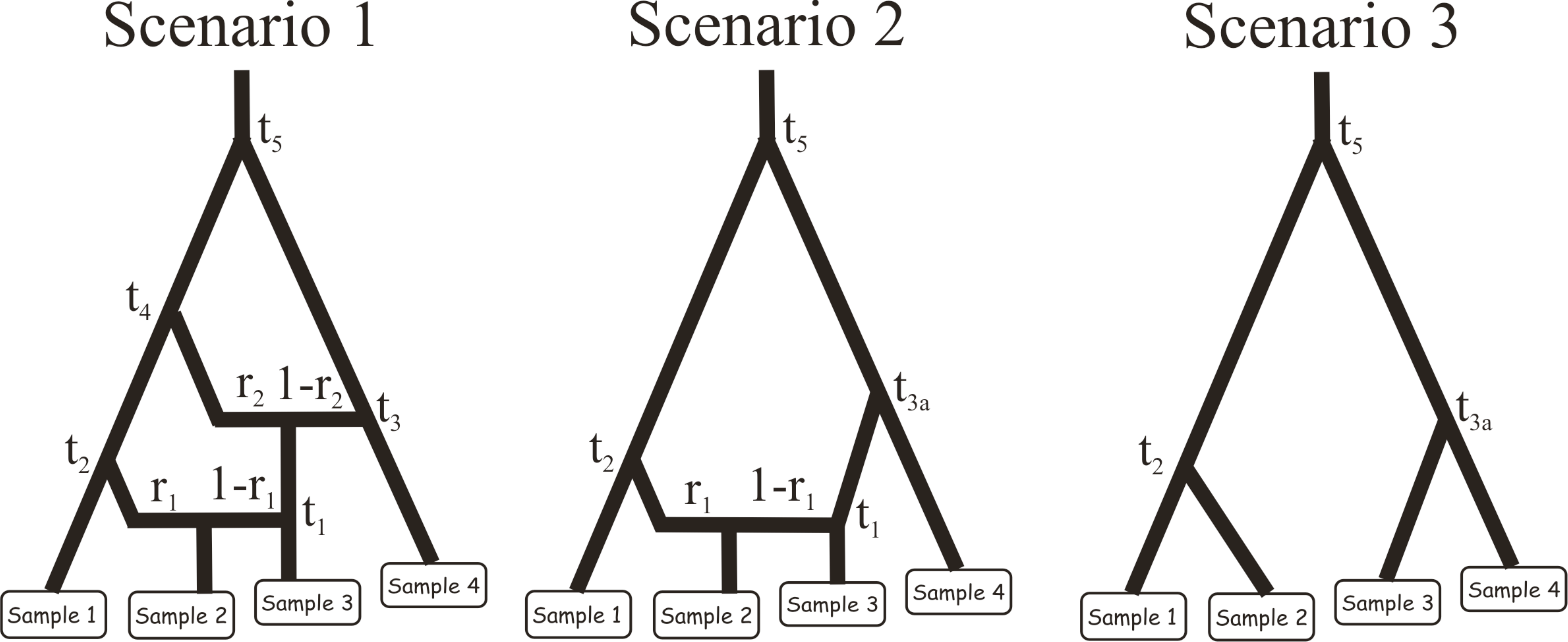}}
\caption{First example: the three evolutionary scenarios. The data set used as an example has been simulated according to scenario 1 (left). The parameter values were the following: all populations had an effective (diploid) size of 1,000, the times of successive events (backward in time) were $t_1$=10, $t_2$=500, $t_3$=10,000, $t_4$=20,000 and $t_5$=200,000, the two admixture rates were $r_1$=0.6 and $r_2$=0.4. Scenario 1 includes 6 populations, the four that have been sampled and two \emph{left} parental populations in the admixture events. Scenario 2 and 3 include 5 and 4 populations respectively. Samples 3 and 4 have been collected 2 and 4 generations earlier than the first two samples, hence their slightly upward locations on the graphs. Time is not at scale.}
\end{figure} 

To answer the first question,  the $n_\delta$=60,000 (1\%) simulated data sets closest to the pseudo-observed data set are selected for the logistic regression and $n_\delta$=600 (0.01\%) for the direct approach. The answer appears in two graphs (upper row in Figure  S7). Both approaches are congruent and show that scenario 1 is significantly better supported by data than any other scenarios. \

To answer the second question,  scenario 1 is chosen and posterior distributions of parameters are estimated taking the 20,000 (1\%) closest simulated data sets, after applying a \emph{logit} transformation of parameter values. Here again, the output is mostly graphical. Each graph provides the prior and posterior distributions of the corresponding parameter (Figure S8). Below each graph are given the mean, median and mode as well as four quantiles (0.025, 0.05, 0.95 and 0.975) of the posterior distribution (Table S1 in Supplementary material). Since the true values are known, we can remark that some parameters are rather well estimated with peaked posteriors such as the common effective population size and the two admixture rates whilst other including all time parameters suggest that data are not very informative for them. Very similar results (not shown) have been obtained with 5,000 and 40,000 simulated data sets selected for the local linear regression, as well as when using a smaller reference table (1 million data sets).\

To evaluate the confidence that can be put into the posterior probability of scenarios, 500 test data sets were simulated with each scenario and known parameter values (i.e. the same values as those used to produce the original data set). Posterior probabilities of the three scenarios were estimated as above and used to compute type I and II errors in the choice of scenario. Results show that scenario 3 is always rightly chosen or excluded. Consequently type I error for scenario 1 is identical to type II error for scenario 2 and vice-versa. For scenario 1, type I errors amount to 0.414 and 0.3 for the direct approach and the logistic regression respectively whereas type II errors amount to 0.014 and 0.020 (cf Figures S9, S10 and S11 for detailed distributions of scenario probabilities). The 500 test data sets simulated with scenario 1 have also been used to estimate posterior distributions of parameters, taking the same proportion (1\%) of closest simulated data sets as above. Relative biases and dispersion measures are given in Table S2 (upper part).  It is clear that several parameters are biased and/or dispersed, the worst case being that of parameter $t_1$. The bias is undoubtedly related to the lack of information in the data, so that point estimates are drawn towards the mean values of prior distributions. The effect of prior distribution is also illustrated in the lower part of Table S2 which provides the same measures, but obtained with different prior distributions for effective size and time of event parameters.

\subsubsection{Illustration on a real data set} 
  
Our second example concerns populations of the  Silvereye, \emph{Zosterops lateralis lateralis}  \citep{EC2003}. During the 19th and 20th century, this bird colonised Southwest Pacific islands from Tasmania. The importance of serial founder events in the microevolution of this species has been questioned in a study based on a six microsatellite loci data set \citep{CD2002}.\ 

\begin{figure}[h]
\centerline{\includegraphics[scale=0.47]{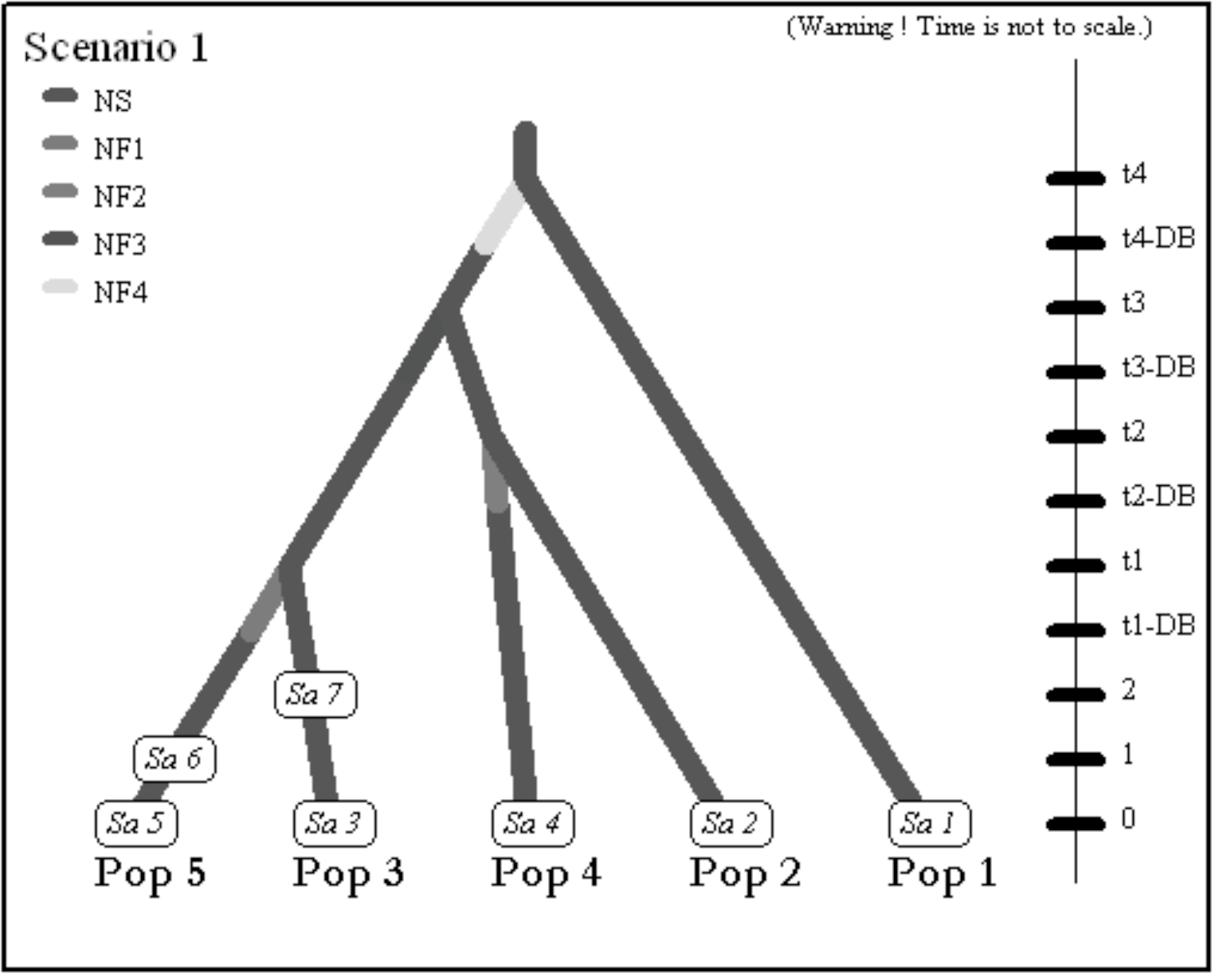}}
\caption{Second example: screenshot of the scenario used in the analysis of the \emph{Zosterops lateralis lateralis} data set. 
In 1830,\emph{ Z. l. lateralis} colonised the South Island of New Zealand (Pop2) from Tasmania (Pop1). In the following years, the population began expanding and dispersing, and reached the North Island by 1856 (Pop3). Chatham Island (Pop4) was colonised in 1856 from the South Island, and Norfolk Island (Pop5) was colonised in 1904 from the North Island (historical information reviewed in Estoup and Clegg 2003). Sample collection times are 1997 for Tasmania (Sa1), South and North island of New Zealand (Sa2 and Sa3, respectively), Chatham island (Sa4) and Norfolk island (Sa5), 1994 for the second sample from Norfolk (Sa6), and 1992 for the second sample from the North island of New Zealand (Sa7). Splitting events and sampling dates in years were translated in number of generations since the most recent sampling date by assuming a generation time of three years (Estoup and Clegg 2003). We hence fixed t1, t2, t3 and t4 to 31, 47, 47 and 56 generations, respectively.}
\end{figure}

Our analysis with $DIYABC$ differs by at least four aspects from the initial ABC analysis processed from the same data set by \citet{EC2003}. First all island populations are treated here in the same analysis whereas, for tractability reasons, independent analyses were made using  all pairs of populations. Second, the initial treatment was based on the algorithm  of \citet{P1999}, whereas $DIYABC$ uses the local linear regression method of \citet{B2002}, which allows a larger number of statistics (see below) and hence makes a better use of data. Third, we have chosen here non-informative flat priors for all demographic parameters. Fourth, because $DIYABC$ is able to treat samples collected at different times, we did not have to pool samples collected at different years from the same island and average sample year collection over islands. We hence end up with a colonization scenario involving five populations and seven samples, two samples having been collected at different times in two different islands (Figure 2). The sequence and dates of colonisation by silvereyes to New Zealand (South and North Island) and Chatham and Norfolk islands have been historically documented. This allows the times for the putative population size fluctuation events in the coalescent gene trees to be fixed, thus limiting the number of parameters. Our scenario was specified by six unknown demographic parameters: the stable effective population size ($N_S$) and the duration of the initial bottleneck ($D_B$), both  assumed to be the same in all islands and  potentially different effective number of founders in Norfolk, Chatham and  South and North island of New Zealand ($N_{F1}$, $N_{F2}$, $N_{F3}$ and $N_{F4}$, respectively). As in \citet{EC2003}, we also assumed that all populations evolved as totally isolated demes after the date of colonisation.\\
We chose uniform priors $U[300, 30000]$ for $N_S$, $U[2, 500]$ for all $N_{Fi}$ and $U[1, 5]$ for $D_B$.
  Prior information regarding the mutation rate and model for microsatellites was the same as in the previous example. Summary statistics included the mean number of alleles, the mean genic diversity \citep{N1987}, the mean coefficient $M$ \citep{GW2001}, $F$st between pairs of population samples \citep{WC1984}, and the \emph{mean classification index}, also called \emph{mean individual assignment likelihood} \citep{PC2007}. \
We produced a reference table with 1 million simulated data sets and estimated parameter posterior distributions taking the 10,000 (1\%) simulated data sets closest to the observed data set for the local linear regression, after applying a $logit$ transformation to parameter values. Similar results were obtained when taking the 2,000 to 20,000 closest simulated data sets and when using a $log$ or $log-tangent$ transformation of parameters as proposed in \citet{EB2004} and \citet{HS2005} (options available in $DIYABC$).\\
\begin{table}
\begin{tabular}[t]{|l|c|c|c|c|c|c|c|}
\hline
Parameter & mean & median & mode &  $Q_{0.050}$ & $Q_{0.950}$ & st. deviation\\
\hline
$N_S$  & 9,399  & 7,446 & 4,107 &  2,706 & 23,007  & 6,273\\
$N_{F1}$  & 19     & 18       & 16       & 9         & 33 & 8.7\\
$N_{F2}$  & 202   & 173    & 108     & 55    & 435 & 118\\
$N_{F3}$  & 197     & 168    & 112     & 55       & 430 & 116\\
$N_{F4}$  & 293   & 288    & 278     & 129     & 470 & 105\\
\hline
\end{tabular} 
\caption{Second example : mean, median, mode, quantiles and standard deviation of posterior distribution samples for effective population sizes (original parameters)  for  the \emph{Zosterops lateralis lateralis} data set.}
\end{table}

\begin{figure}[h]
\centerline{\includegraphics[scale=0.7]{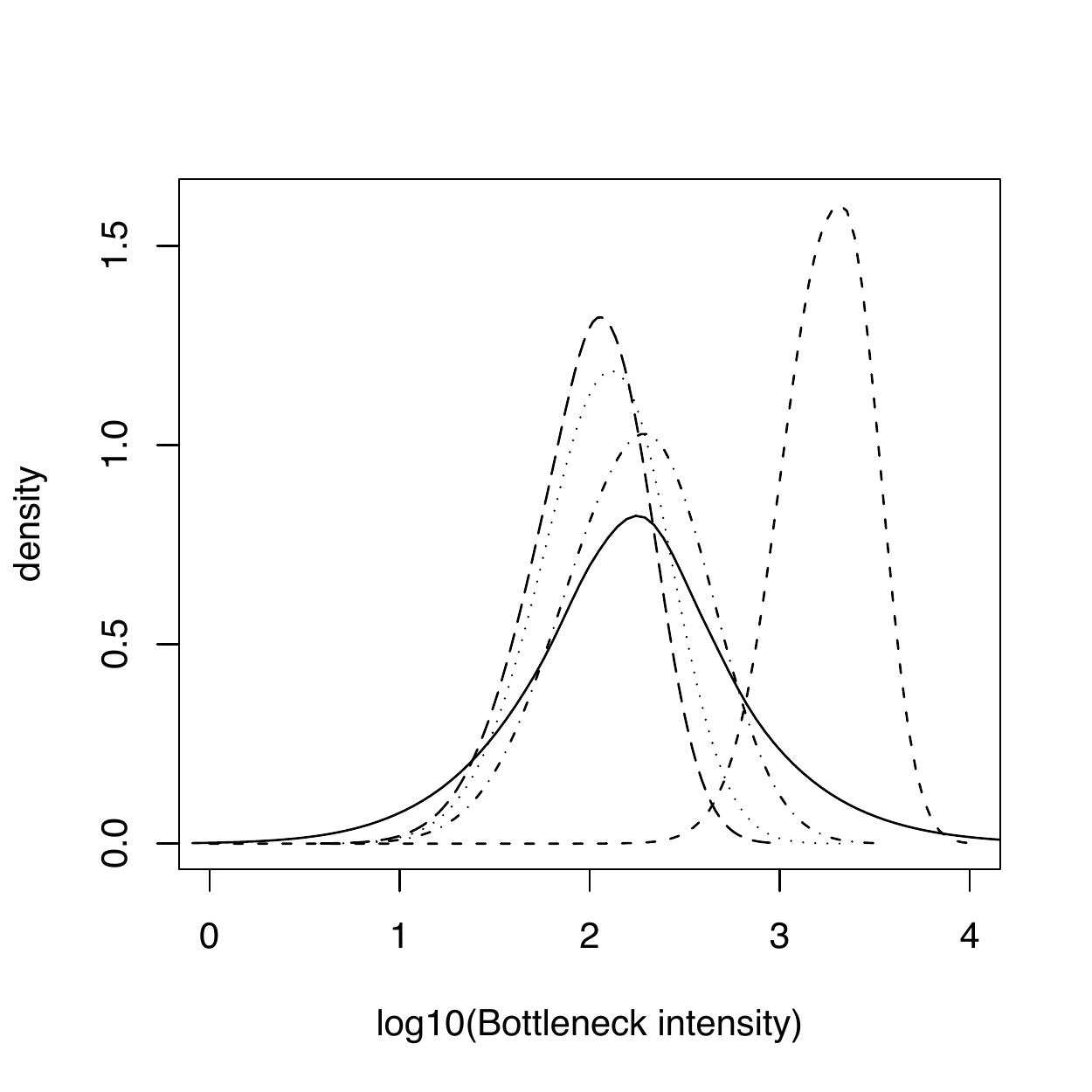}}
\caption{Second example: posterior distributions of the bottleneck severity (see definition in text) for the invasions of four Pacific islands by \emph{Zosterops lateralis lateralis}. The four discontinuous lines with small dashes, dots, dash-dots and long dashes correspond to Norfolk, Chatham, North Island and South Island of New Zealand, respectively. The continuous line corresponds to the prior distribution, which is identical for each island. This graph has been made with the \emph{locfit} function of the R statistical package \citep{IG1996}, using an option of $DIYABC$ which saves the sample of the parameter values adjusted by the local linear regression \citep{B2002}. }
\end{figure}

Results for the main demographic parameters are presented in Table 1. They indicate the colonization by a small number of founders and/or a slow demographic recovery after foundation for Norfolk island only (median $N_{F1}$ value of 18 individuals). Other island populations appear to have been founded by silvereye flocks of larger size and/or have recovered quickly after foundation. In agreement with this, the bottleneck severity (computed as $BS_i$=$D_B \times N_S/N_{Fi}$) was more than one order of magnitude larger for the population from Norfolk than for other island populations (Figure 3). These results are in the same vein as those obtained by \citet{EC2003} and agree with their main conclusions. Discrepancies in parameter estimation are observed however (e.g. larger $N_S$ values and more precise inferences for $N_{F2}$, $N_{F3}$ and $N_{F4}$ in the present treatment). This was expected due to the differences in the methodological design underlined above. With the possibility of treating all population samples together, $DIYABC$ allows a more elaborate and satisfactory treatment compared to previous analyses \citep{EC2003,ME2005}.

\section{Conclusion} So far, the ABC approach has remained inaccessible to most biologists because of the complex computations involved. With $DIYABC$, non specialists can now perform ABC-based inference on various and complex population evolutionary scenarios, without reducing them to simple standard situations, and hence making a better use of their data. In addition, this programs also allows them to compare competing scenarios and quantify their relative support by the data. Eventually, it provides a way to evaluate the amount of confidence that can be put into the various estimations.
The main limitations of the current version of $DIYABC$ are the assumed absence of migration among populations after they have diverged and the mutation models which mostly refer to microsatellite loci. Next developments will aim at progressively removing these limitations.\\
\section*{Funding}
The development of $DIYABC$ has been supported by a grant from the French Research National Agency (project $MISGEPOP$) and an EU grant awarded to JMC as an EIF Marie-Curie Fellowship (project $StatInfPopGen$) that allowed him to spend two years in D.J.B.'s Epidemiology and Public Health department at Imperial College (London, UK) where he wrote a major part of this program.
\section*{Acknowledgements}
We are most grateful to Sonya Clegg for authorizing the use of Silvereye microsatellite data in our second example.

\end{document}